\documentclass[aps,pra,twocolumn,groupedaddress,floatfix]{revtex4}

\usepackage{graphicx}
\usepackage{textcomp}

\begin{document}

\title{~\vspace{1.5cm}\\ Double-resonant extremely asymmetrical scattering
of electromagnetic waves in periodic arrays separated by a gap}


\author{D. K. Gramotnev}
%\thanks{}

\author{T. A. Nieminen}
%\email[]{timo@physics.uq.edu.au}

\affiliation{Centre for Medical and Health Physics,
School of Physical Sciences, Queensland University of Technology,
GPO Box 2434, Brisbane, QLD 4001, Australia}


\date{28th October 1999}

\begin{abstract}
\vspace{-6cm}
\noindent
\hspace{-1.5cm}\textbf{Preprint of:}

\noindent
\hspace{-1.5cm}D. K. Gramotnev and T. A. Nieminen,

\noindent
\hspace{-1.5cm}``Double-resonant extremely asymmetrical scattering of

\noindent
\hspace{-1.5cm}electromagnetic waves in periodic arrays separated by a gap'',

\noindent
\hspace{-1.5cm}\textit{Optical and Quantum Electronics} \textbf{33},
1--17 (2001)

\hrulefill

\vspace{4cm}

Two strong simultaneous resonances of scattering---double-resonant
extremely asymmetrical scattering (DEAS)---are predicted in two parallel,
oblique, periodic Bragg arrays separated by a gap, when the scattered wave
propagates parallel to the arrays. One of these resonances is with respect
to frequency (which is common to all types of Bragg scattering), and another
is with respect to phase variation between the arrays. The diffractional
divergence of the scattered wave is shown to be the main physical reason
for DEAS in the considered structure. Although the arrays are separated,
they are shown to interact by means of the diffractional divergence of the
scattered wave across the gap from one array into the other. It is also
shown that increasing separation between the two arrays results in a broader
and weaker resonance with respect to phase shift. The analysis is based on a
recently developed new approach allowing for the diffractional divergence of
the scattered wave inside and outside the arrays. Physical interpretations
of the predicted features of DEAS in separated arrays are also presented.
Applicability conditions for the developed theory are derived.
\end{abstract}

%\pacs{42.62.Be,42.62.Eh,42.25.Fx,42.25.Ja}
%\keywords{}

\maketitle

\section{Introduction}

Extremely asymmetrical scattering (EAS) occurs when the scattered wave
propagates parallel to the front boundary of a strip-like periodic Bragg
array~\cite{kishino1971,kishino1972,bedynska1973,bedynska1974,%
bakhturin1995,gramotnev1995,gramotnev1996,gramotnev1997,gramotnev1999a,%
gramotnev1999b,gramotnev1999c}. Steady-state EAS is characterized by a
strong resonant increase of the scattered wave amplitudes inside and
outside the array. The smaller the grating amplitude, the larger the
amplitudes of the scattered waves~\cite{kishino1972,bakhturin1995,%
gramotnev1995,gramotnev1996,gramotnev1997,gramotnev1999a,gramotnev1999b,%
gramotnev1999c}. In addition, the incident and scattered waves inside the
array split into three waves each~\cite{bakhturin1995,gramotnev1995,%
gramotnev1996,gramotnev1997,gramotnev1999a,gramotnev1999b,gramotnev1999c}.
Two of these scattered waves and two of the incident waves inside the array
are evanescent waves localized near the array
boundaries~\cite{bakhturin1995,gramotnev1995,gramotnev1996,gramotnev1997}.
The third scattered wave is a plane wave propagating at a grazing angle into
the array~\cite{bakhturin1995,gramotnev1995,gramotnev1996,gramotnev1997}.
All these features demonstrate that EAS is radically different from the
conventional Bragg scattering in periodic arrays.

There are two opposing physical mechanisms affecting scattering in the
extremely asymmetric geometry~\cite{bakhturin1995,gramotnev1995,%
gramotnev1996,gramotnev1997,gramotnev1999a,gramotnev1999b,gramotnev1999c}.
On the one hand, the scattered wave amplitude must increase along the
direction of its propagation (parallel to the front boundary of the
periodic array) due to scattering of the incident wave inside the array.
On the other hand, the scattered wave amplitude must decrease along the
direction of its propagation due to the diffractional divergence of this
wave~\cite{bakhturin1995,gramotnev1995,%
gramotnev1996,gramotnev1997,gramotnev1999a,gramotnev1999b,gramotnev1999c}.
This diffractional divergence was demonstrated to be the main physical reason
for EAS. In the steady-state case of scattering in the geometry of EAS,
the contributions to the scattered wave amplitude from the two opposing
mechanisms must exactly compensate each other. A new powerful approach for
a simple theoretical analysis of EAS, based on allowance for the diffractional
divergence of the scattered wave, was introduced and
justified~\cite{bakhturin1995,gramotnev1995,%
gramotnev1996,gramotnev1997,gramotnev1999a,gramotnev1999b,gramotnev1999c}.

It was demonstrated that in non-uniform arrays with varying phase of the
grating, EAS is characterized by two simultaneous resonances---one with
respect to frequency, and the other with respect to the phase variation
in the grating~\cite{gramotnev1999a,gramotnev1999c,gramotnev2000}.
As a result, typical scattered wave amplitudes in this
case appear to be many times larger than those for EAS in uniform arrays.
This effect was called double-resonant extremely asymmetrical scattering
(DEAS)~\cite{gramotnev1999a,gramotnev1999c,gramotnev2000}. DEAS was
described for a non-uniform array that consists of two joint strip-like
periodic arrays with different phases of the grating, i.e. a step-like
variation of the grating phase occurs at the interface between the
arrays~\cite{gramotnev1999a,gramotnev1999c,gramotnev2000}.

The main physical reason for DEAS is related to the diffractional divergence
of the scattered wave from one of the joint periodic arrays into another.
For example, the scattered wave from the second array, propagating parallel
to the array boundaries, diverges into the first array, and is re-scattered
by the grating in the first array. Due to the phase difference between the
arrays (that should be close to $\pi$), the resultant re-scattered wave
appears to be approximately in phase with the incident wave inside the first
array. Therefore, the amplitude of the incident wave (that is analogous to a
force driving resonant oscillations) is increased due to the constructive
interference with the mentioned re-scattered wave. The same speculations
are valid for the second array. This is the reason for a substantial
resonant increase in the scattered wave amplitude in DEAS compared to
EAS~\cite{gramotnev1999a,gramotnev1999c,gramotnev2000}.

It was also demonstrated that strong DEAS (i.e. strong resonance with
respect to phase variation between the arrays) can take place only if
the array width is smaller than a critical width. Physically, half of
this critical width is equal to a distance through which the scattered
wave can be spread inside the array due to the diffractional divergence
before being re-scattered in the
grating~\cite{gramotnev1999a,gramotnev1999c,gramotnev2000}.
It is obvious that only within this distance from the interface between
the two joint arrays can the diffractional divergence significantly affect
scattering.

The aim of this paper is to demonstrate that strong DEAS can take place
not only in two joint periodic arrays, but also in two oblique, strip-like,
periodic arrays separated by a gap. We will show that separated arrays can
interact with each other by means of the diffractional divergence of the
scattered waves across the gap. This interaction will be demonstrated to
decrease with increasing gap width. The effect of gap width on the incident
and scattered wave amplitudes inside and outside the arrays will be
investigated. Steady-state DEAS of bulk and guided optical waves will
be analyzed. Applicability conditions for the developed approximate
theory will be determined and investigated.

\section{Coupled wave equations}

In this section we present coupled wave equations and their solutions
in the case of DEAS of bulk TE electromagnetic waves in two separated
uniform periodic arrays (Fig. 1) that are represented by small sinusoidal
variations of the mean dielectric permittivity:
\begin{equation}
\begin{array}{l}
\epsilon_s = \epsilon + \epsilon_1
\exp(\mathrm{i}\mathbf{q}\cdot\mathbf{r})
+ \epsilon_1^\ast \exp(-\mathrm{i}\mathbf{q}\cdot\mathbf{r}),\\
\epsilon_s = \epsilon + \epsilon_2
\exp(\mathrm{i}\mathbf{q}\cdot\mathbf{r})
+ \epsilon_2^\ast \exp(-\mathrm{i}\mathbf{q}\cdot\mathbf{r}),\\
\\
\epsilon_s = \epsilon,
\end{array}
\begin{array}{l}
\textrm{if}\;\; 0 < x < L_1;\\
\textrm{if}\;\; L+L_1 < x\\ < L+L_1+L_2;\\
\textrm{otherwise,}
\end{array}
\end{equation}
where $L_1$ and $L_2$ are the widths of the first and second arrays
(Fig. 1), $L$ is the width of the gap between the arrays, $\mathbf{q}$
is the reciprocal lattice vector, $|\mathbf{q}| = 2\pi/\Lambda$, $\Lambda$
is the grating period that is the same in both the arrays, the mean
dielectric permittivity $\epsilon$ is the same in all parts of the
structure (inside and outside the arrays), and the amplitudes of the
gratings $\epsilon_1$ and $\epsilon_2$ in the first and the second
arrays are assumed to be small:
\begin{equation}
|\epsilon_{1,2}|/\epsilon \ll 1.
\end{equation}
The grating amplitudes $\epsilon_1$ and $\epsilon_2$ can differ in
magnitude and/or in phase. There is no dissipation of electromagnetic
waves inside or outside the array, i.e. $\epsilon$ is real and positive.

\begin{figure}[t]
\includegraphics[width=\columnwidth]{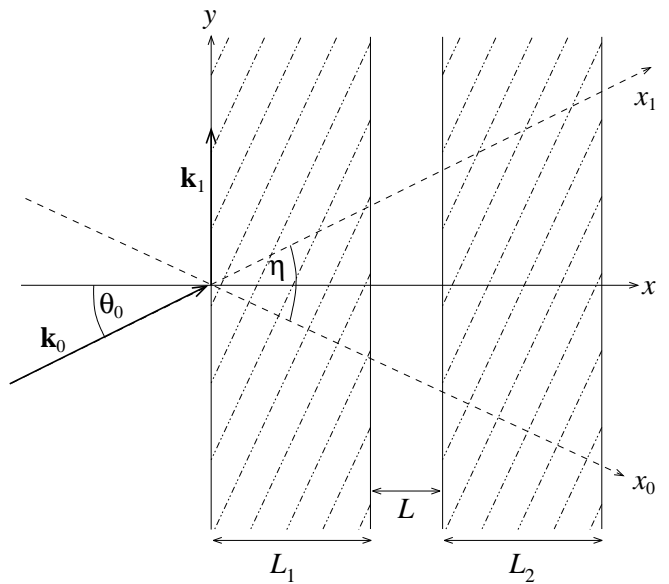}
\caption{The geometry of DEAS in two non-uniform strip-like periodic
Bragg arrays of width $L_1$ and $L_2$, separated by a gap of width $L$.}
\end{figure}

A plane TE electromagnetic wave (with the electric field parallel to the
$z$-axis) is incident onto the first array at an angle $\theta_0$
(measured from the $x$-axis counter-clockwise---Fig. 1). We assume that
the Bragg condition is satisfied precisely:
\begin{equation}
\mathbf{k}_1 - \mathbf{k}_0 = -p\mathbf{q},
\end{equation}
where $p$ takes one of the values: $\pm 1, \pm 2, ...$, $\mathbf{k}_0$
is the wave vector of the incident wave, $\mathbf{k}_1$ is parallel to
the array boundaries (Fig. 1),
$|\mathbf{k}_1| = |\mathbf{k}_0| = k_0 = \omega \epsilon^{1/2}/c$,
$\omega$ is the angular frequency, and $c$ is the speed of light in vacuum.

As has been demonstrated previously~\cite{bakhturin1995,gramotnev1995,%
gramotnev1996,gramotnev1997,gramotnev1999a,gramotnev1999b,%
gramotnev1999c,gramotnev2000}, strong resonant increase in the scattered
wave amplitude in the extremely asymmetrical geometry can only occur for
small grating amplitudes---see condition (2). In this case, the two wave
approximation~\cite{kogelnik1969} is valid, and only two harmonics in the
Floquet expansion---incident and scattered waves---need to be taken into
account inside and outside the array:
\begin{eqnarray}
E(x) & = & E_0(x)\exp\left\{ \mathrm{i}k_{0x}x + \mathrm{i}k_{0y}y
- \mathrm{i}\omega t \right\}\nonumber \\ & & + 
E_1(x)\exp\left\{ \mathrm{i}k_0 y - \mathrm{i}\omega t \right\},
\end{eqnarray}
where $E_0(x)$ and $E_1(x)$ are the varying amplitudes of the electric
fields in the incident and scattered waves, respectively,
$k_{0x} = k_0 \cos\theta_0$, and $k_{0y} = k_0 \sin\theta_0$.
Detailed discussions of applicability conditions for this approximation
in the cases of EAS and DEAS are presented below in Section 4.

Similarly to how it was done in~\cite{bakhturin1995,gramotnev1996,%
gramotnev2000,gramotnev1999a,gramotnev1999c}, the coupled wave equations
in the separated arrays, describing DEAS, can be derived by means of
separate analyses of the diffractional divergence of the scattered wave
(by means of the parabolic equation of diffraction~\cite{bakhturin1995,%
gramotnev1999a,gramotnev1999c,gramotnev2000} or Fourier
analysis~\cite{gramotnev1996,gramotnev2000}), and scattering (by means
of the conventional theory of scattering~\cite{kogelnik1969,%
gramotnev1996,stegeman1981,popov1985,wellerbrophy1988,hall1990}).
In the steady-state DEAS the contribution to the scattered wave amplitude
from scattering is exactly compensated by the contribution from the
diffractional divergence. In the same fashion as
in~\cite{gramotnev1996,gramotnev2000,gramotnev1999c}, the comparison of
these contributions leads to the coupled wave equations in the separated
arrays:
\begin{equation}
\frac{\mathrm{d}}{\mathrm{d}x} E_0(x) = \left\{
\begin{array}{l}
\mathrm{i} K_{11} E_1(x)\\
0\\
\mathrm{i} K_{12} E_1(x)\\
~
\end{array}
\begin{array}{l}
\textrm{if}\;\; 0 \le x \le L_1,\\
\textrm{if}\;\; L_1 < x < L+L_1,\\
\textrm{if}\;\; L+L_1 \le x\\ \le L+L_1+L_2,
\end{array}
\right.
\end{equation}
\begin{equation}
\frac{\mathrm{d}^2}{\mathrm{d}x^2} E_1(x) = \left\{
\begin{array}{l}
\mathrm{i} -K_{01} E_0(x)\\
0\\
\mathrm{i} K_{02} E_0(x)\\
~
\end{array}
\begin{array}{l}
\textrm{if}\;\; 0 \le x \le L_1,\\
\textrm{if}\;\; L_1 < x < L+L_1,\\
\textrm{if}\;\; L+L_1 \le x\\ \le L+L_1+L_2,
\end{array}
\right.
\end{equation}
where
\begin{equation}
K_{1j} = \Gamma_{1j}\cos\eta/\cos\theta_0,
\end{equation}
\begin{equation}
K_{0j} = -2k_1\Gamma_{0j}\sin(\eta-\theta_0),
\end{equation}
indices $j = 1,2$ correspond to the first and second arrays, respectively,
$\eta$ is the angle between the direction of propagation of the incident
wave and the $x_0$-axis (Fig. 1), $\Gamma_{0j}$ and $\Gamma_{1j}$
are the coupling coefficients in the well-known coupled wave
equations~\cite{kogelnik1969,stegeman1981,popov1985,wellerbrophy1988,%
hall1990,gramotnev1996}
\begin{equation}
\mathrm{d}E_0/\mathrm{d}x_0 = \mathrm{i}\Gamma_{1j}E_1,\;\;\;
\mathrm{d}E_1/\mathrm{d}x_0 = \mathrm{i}\Gamma_{0j}E_0
\end{equation}
for the conventional dynamic theory of scattering in two isolated
uniform arrays with the grating fringes parallel to the array boundaries,
and with the grating amplitudes $\epsilon_1$ (for $j = 1$) and
$\epsilon_2$ (for $j = 2$).

Note that Equations (5)--(9) are directly applicable for the description
of DEAS of all types of waves, including optical modes guided by a slab
with a periodically corrugated boundary. In this case, the plane of Fig. 1
is the plane of a guiding slab, and the incident and scattered waves are
modes guided by this slab. The only difference between scattering of
different types of waves are different values of coupling coefficients
$\Gamma_{0j}$ and $\Gamma_{1j}$ that are already determined in the
conventional theory of scattering~\cite{kogelnik1969,stegeman1981,%
popov1985,wellerbrophy1988,hall1990,gramotnev1996}. For example, in
the case of bulk TE electromagnetic waves in arrays described by
Equation (1) we obtain~\cite{kogelnik1969,gramotnev1996}:
\begin{equation}
\Gamma_{0j} = - \Gamma_{1j}^\ast = -\epsilon_j^\ast\omega^2/[2c^2k_0\cos\eta]. 
\end{equation}
We have also used $k_1$ instead of $k_0$ in Equation (8), because for
guided optical modes $k_1$ may not be equal to $k_0$ (e.g., for scattering
of TE modes guided by a slab into TM modes of the same slab).

The solutions to coupled wave equations (5) and (6) inside the first
($j = 1$) and the second ($j = 2$) arrays can be written
as~\cite{bakhturin1995,gramotnev1996,gramotnev1999b,gramotnev2000}:
\begin{eqnarray}
E_1(x) &=& C_{1j}\exp(\mathrm{i}\lambda_{1j}x)
+ C_{2j}\exp(\mathrm{i}\lambda_{2j}x)\nonumber \\& &
+ C_{3j}\exp(\mathrm{i}\lambda_{3j}x),\nonumber \\
E_0(x) &=& D_{1j}\exp(\mathrm{i}\lambda_{1j}x)
+ D_{2j}\exp(\mathrm{i}\lambda_{2j}x) \\ & &
+ D_{3j}\exp(\mathrm{i}\lambda_{3j}x)\nonumber,
\end{eqnarray}
where $\lambda_{1j} = \gamma_j$,
$\lambda_{2j} = - (1-\mathrm{i}3^{1/2})\gamma_j/2$,
$\lambda_{2j} = - (1+\mathrm{i}3^{1/2})\gamma_j/2$, and
\begin{equation}
\gamma_j = [K_{0j}K_{1j}]^{1/3}.
\end{equation}
In front of and behind the arrays,
\begin{equation}
\begin{array}{l}
E_1(x) = A_1,\\
E_1(x) = A_3,
\end{array}
\begin{array}{l}
E_0(x) = E_{00},\\
E_0(x) = E_{01},
\end{array}
\begin{array}{l}
\textrm{if}\;\; x<0;\\
\textrm{if}\;\; x>L+L_1+L_2.
\end{array}
\end{equation}
In the gap between the arrays (i.e. for $L_1 < x < L + L_1$) Equations (5)
and (6) yield:
\begin{equation}
E_1(x) = A_{21} + A_{22}x, E_0(x) = B.
\end{equation}
These equations suggest that, although there is no scattering in the gap
between the arrays, the scattered wave amplitude is not constant due to
the interaction between the scattered waves diverging from each of the two
arrays.

A relationship between the amplitudes of the incident and scattered waves
inside the array $C_{ij}$ and $D_{ij}$ ($j = 1,2$; $i = 1,2,3$) can be
established by substituting Equations (11) into Equations (6):
\begin{equation}
C_{ij} = \lambda_{ij}D_{ij}/K_1
\end{equation}

The remaining unknown constants $A_1$, $A_3$, $A_{21}$, $A_{22}$, $B$,
$E_{01}$, and $D_{ij}$ can be determined from the boundary
conditions~\cite{gramotnev1999a,gramotnev1999c} of continuity of the fields
and their derivatives across the four array boundaries (note however, that
in the considered two-wave approximation, the derivatives of the fields in
the incident wave are not continuous across the
boundaries~\cite{gramotnev1996,gramotnev1997,gramotnev1999a,gramotnev1999b,%
gramotnev1999c,gramotnev2000}. The resultant analytical equations for these
constants (amplitudes) appear to be too awkward to be presented here.
Instead, in the next section we analyze these solutions numerically for
different gap widths and array parameters.

\section{Analysis of DEAS in separated arrays}

As expected, the form of solutions (11)--(13) is the same as for two
joint uniform arrays~\cite{gramotnev1999a,gramotnev2000}. However, due to
the presence of the gap with the field given by Equation (14), the overall
pattern of scattering changes substantially. That is, the particular
dependencies of the incident and scattered wave amplitudes $E_1(x)$ and
$E_0(x)$ inside and outside the arrays are noticeably different from those
obtained for EAS and DEAS in uniform and non-uniform arrays without the
gap~\cite{gramotnev1999a,gramotnev1999b,gramotnev1999c,gramotnev2000}.
This will be seen from the figures below.

Equations (11)--(15) are also written in such a way that they are valid
for any types of waves, including guided TE and TM modes in a slab with
a corrugated interface. The boundary conditions at the array boundaries
will also be the same for all types of waves, because actually there are
no physical boundaries since the mean parameters of the media are the same
inside and outside the array. Therefore, we can consider bulk TE
electromagnetic waves in arrays described by Equation (1), keeping in
mind that all the results below are valid, for example, for DEAS of
guided optical modes~\cite{gramotnev1999a,gramotnev1999c,gramotnev2000}.

\begin{figure}[htb]
\includegraphics[width=\columnwidth]{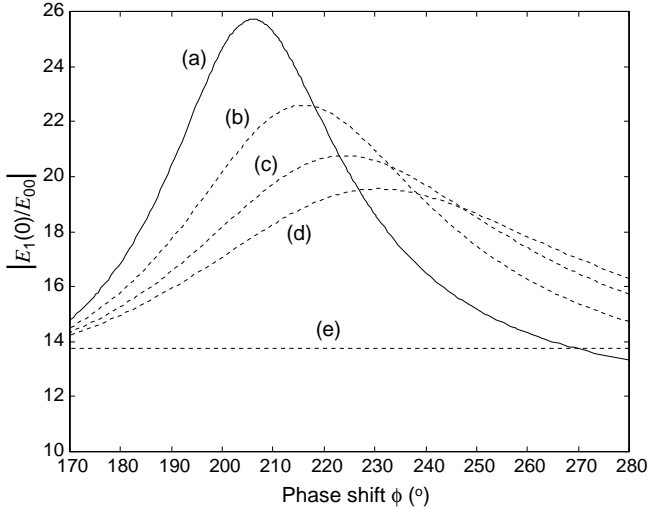}
\caption{The dependencies of relative (normalized) amplitudes of the
scattered wave at the front array boundary on phase shift $\phi$ between
the arrays for different values of gap width $L$: (a) $L = 0$,
(b) $L = 3$\,{\textmu}m, (c) $L = 6$\,{\textmu}m, (d) $L = 9$\,{\textmu}m,
and (e) $L = +\infty$. Scattering of bulk TE electromagnetic waves takes
place in the structure with $\epsilon = 5$, $\epsilon_1 = |\epsilon_2|
= 5\times 10^{-3}$, $\epsilon_2 = \epsilon_1 \exp(\mathrm{i}\phi)$,
$\theta_0 = \pi/4$, $L_1 = L_2 = 15$\,{\textmu}m,
$\lambda_\mathrm{vacuum} = 1$\,{\textmu}m, the grating period
$\Lambda \approx 0.58$\,{\textmu}m, $\eta = 3\pi/8$.}
\end{figure}

Fig. 2 presents typical dependencies of the relative scattered wave
amplitude at the front boundary of the first array (i.e. at $x = 0$)
on the phase variation $\phi$ between the arrays for different values
of the gap width $L$. The structural parameters are as follows:
$\epsilon =5$, $\epsilon_1 = |\epsilon_2| = 5\times 10^{-3}$,
$\epsilon_2 = \epsilon_1 \exp(\mathrm{i}\phi)$,
$\theta_0 = \pi/4$, $L_1 = L_2 = 15$\,{\textmu}m, i.e. the two separated
arrays, apart from a phase difference $\phi$ between them, are identical.
The period of the structure $\Lambda$ is also the same for both the arrays,
and is unambiguously determined by the Bragg condition (3) and the angle of
incidence $\theta_0$: $\Lambda \approx 0.58$\,{\textmu}m, and $\eta = 3\pi/8$.
The wavelength in vacuum is $\lambda = 1$\,{\textmu}m.

Fig. 2 demonstrates that as the gap width decreases, the maximum of the
scattered wave amplitude at an optimal (resonant) value of $\phi$ becomes
sharper and stronger until the gap width reaches zero (curve (a)). It can
be seen that even if the arrays are separated, they can effectively interact
across the gap, which results in DEAS. Since the scattered waves in both the
arrays propagate parallel to the array boundaries (Fig. 1), the interaction
(interference) between the waves can only occur due to their diffractional
divergence from one array into another across the gap. This clearly
demonstrates not only a new interesting effect in periodic arrays in the
extremely asymmetrical geometry, but also most explicitly confirms the
unique role of the diffractional divergence in DEAS.

Curves (b)--(d) in Fig. 2 show the transition from DEAS in the joint
arrays (curve (a)) to EAS in the uniform array of $L_1 = 15$\,{\textmu}m
(curve (e)) as the gap width increases from zero to infinity. Even small
gap widths, e.g., $L = 3$\,{\textmu}m (curve (b) in Fig. 2), which is about
seven wavelengths in the structure, result in a noticeable decrease of the
scattered wave amplitude compared to DEAS in the joint arrays. This is
because the effectiveness of the diffractional divergence in spreading
the scattered wave across the gap quickly decreases with increasing gap width.

\begin{figure}[hb]
\includegraphics[width=\columnwidth]{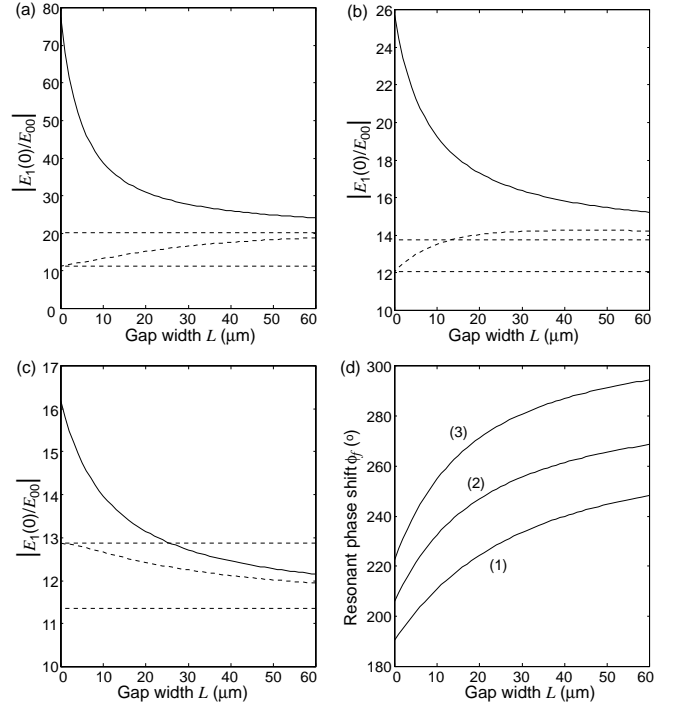}
\caption{Solid and dashed curves in (a)--(c) present the dependencies of
relative scattered wave amplitudes at the front boundary $x = 0$ on gap
width $L$ for bulk TE electromagnetic waves in the structure with
$\epsilon = 5$, $\epsilon_1 = |\epsilon_2| = 5\times 10^{-3}$,
$\epsilon_2 = \epsilon_1 \exp(\mathrm{i}\phi)$, $\theta_0 = \pi/4$,
$\lambda_\mathrm{vacuum} = 1$\,{\textmu}m, $\phi = \phi_{rf}$ (solid curves),
and $\phi = 0$ (dashed curves); (a) $L_1 = L_2 = 10$\,{\textmu}m;
(b) $L_1 = L_2 = 15$\,{\textmu}m; (c) $L_1 = L_2 = 20$\,{\textmu}m.
Dotted lines: $L = +\infty$ (i.e., uniform isolated arrays of widths $L_1$);
dash-and-dot lines: $L = 0$ and $\phi = 0$ (i.e., uniform isolated arrays of
widths $2L_1$). Resonant values $\phi_{rf}$ of the phase shift between
the arrays (i.e., the values of $\phi$ at which the scattered wave
amplitude is maximal at the front boundary $x = 0$) are presented in (d)
as functions of gap width $L$: (1) $L_1 = L_2 = 10$\,{\textmu}m;
(2) $L_1 = L_2 = 15$\,{\textmu}m; (3) $L_1 = L_2 = 20$\,{\textmu}m.}
\end{figure}

The dependencies of maximal scattered wave amplitudes at the front array
boundary (i.e. at $x = 0$) on gap width $L$ in the case of DEAS are
presented by the solid curves in Fig. 3a--c for different array widths:
(a) $L_1 = L_2 = 10$\,{\textmu}m, (b) $L_1 = L_2 = 15$\,{\textmu}m,
(c) $L_1 = L_2 = 20$\,{\textmu}m. Fig. 3d presents the resonant phase
shifts $\phi_{rf}$ between the arrays as functions of gap width $L$
for the same array widths: 10\,{\textmu}m (curve 1), 15\,{\textmu}m
(curve 2), and 20\,{\textmu}m (curve 3). In other words, each value of
$|E_1(0)/E_{00}|$ given by the solid curves in Fig. 3a--c has been
determined at the corresponding resonant value of the phase shift
$\phi = \phi_{rf}$ given by Fig. 3d. The index $f$ in $\phi_{rf}$
indicates that this is the value of the phase shift at which the
scattered wave amplitude is maximal at the front boundary $x = 0$.
Dotted lines in Fig. 3a--c correspond to EAS in the first of the uniform
arrays in the absence of the second array. Obviously, these lines must
also correspond to the case when the gap is infinitely large. Therefore,
if the gap width increases to infinity, the solid curves in Fig. 3a--c
tend to the dotted lines. The dashed curves in Fig. 3a--c present the
amplitudes of the scattered wave at the front boundary of the first
array as functions of gap width $L$ for zero phase shift $\phi = 0$
(i.e. for EAS in separated arrays). In this case, the interaction
between the arrays does not result in a significant increase in the
scattered wave amplitude compared to EAS in the first uniform array
(the dotted lines). The amplitude of the scattered wave at the front
boundary of the first array basically varies from its value for a
uniform array of the width $2L_1$ (dash-and-dot lines in Fig. 3a--c)
to the value for the first isolated uniform array (dashed curves in
Fig. 3a--c). This clearly demonstrates strong differences between DEAS
and EAS in separated arrays. The existence of a phase shift between the
arrays, which is close to the resonant value (Fig. 3d), results in a
substantial resonant increase in the scattered wave amplitude (solid
curves in Fig. 3a--c).

The dependencies of the amplitudes of the incident and scattered waves on
the $x$-coordinate inside and between the arrays is presented in Fig. 4
for resonant values of $\phi = \phi_r$, where $\phi_r$ corresponds to
maximal scattered wave amplitudes in the middle of the gap ($\phi_r$ is
very close to, but not equal to $\phi_{rf}$). It can be seen that the
amplitude of the incident wave in the gap is constant, which is expected
since there is no scattering in the gap, and the diffractional divergence
of the incident wave is negligible (curves 2 and 3 in Fig. 4a, b). A
maximum of the incident wave amplitude is achieved in the gap. However,
this maximum quickly decreases with increasing gap width $L$ (Fig. 4a, b),
reaching $E_{00}$ (the amplitude of the incident wave at the front boundary
$x = 0$) when the gap width is infinite---see curves 4 in Fig. 4a, b).

\begin{figure}[th]
\includegraphics[width=\columnwidth]{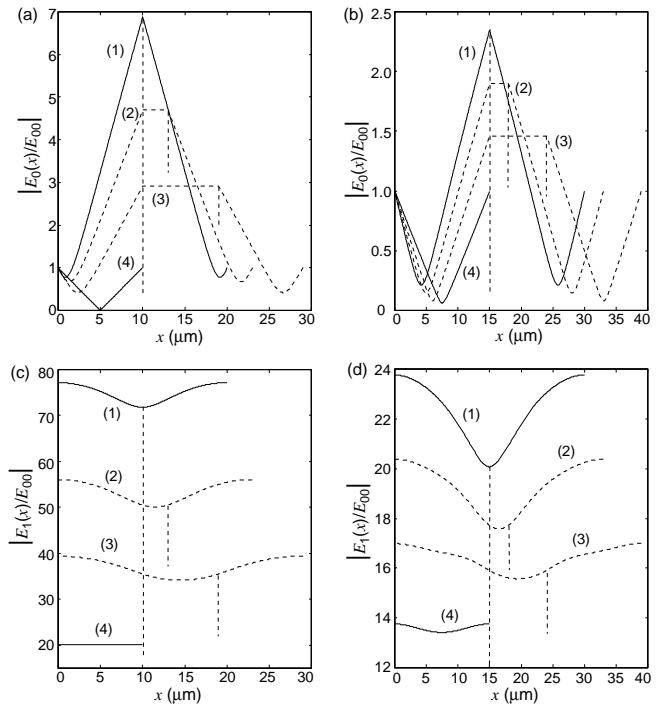}
\caption{Typical dependencies of relative amplitudes of the incident
[(a) and (b)] and scattered [(c) and (d)] waves on distance $x$ from the
front boundary $x = 0$ for narrow ($L_1 = L_2 < L_c$) arrays for different
widths of the gap and arrays: (a) and (c): $L_1 = L_2 = 10$\,{\textmu}m,
$L = 0$ and $\phi = \phi_r \approx 190.3$\textdegree\ (curves 1);
$L = 3$\,{\textmu}m and $\phi = \phi_r \approx 197.2$\textdegree\ (curves 2);
$L = 9$\,{\textmu}m and $\phi = \phi_r \approx 206.6$\textdegree\ (curves 3);
$L = +\infty$ (curves 4). (b) and (d): $L_1 = L_2 = 15$\,{\textmu}m,
$L = 0$ and $\phi = \phi_r \approx 197.6$\textdegree\ (curves 1);
$L = 3$\,{\textmu}m and $\phi = \phi_r \approx 200.9$\textdegree\ (curves 2);
$L = 9$\,{\textmu}m and $\phi = \phi_r \approx 199.3$\textdegree\ (curves 3);
$L = +\infty$ (curves 4). $\phi_r$ is the value of $\phi$ at which the
scattered wave amplitude is maximal in the middle of the gap (note that
in general $\phi_r \ne \phi_{rf}$). All other structural parameters are
the same as for Figs. 3 and 4: $\epsilon = 5$,
$\epsilon_1 = |\epsilon_2| = 5\times 10^{-3}$,
$\epsilon_2 = \epsilon_1 \exp(\mathrm{i}\phi)$, $\theta_0 = \pi/4$,
$\lambda_\mathrm{vacuum} = 1$\,{\textmu}m. The edges of the gap between
the arrays are marked by the vertical dotted lines.}
\end{figure}

Fig. 4c, d demonstrate how the increase in the gap width from zero to
infinity results in the transformation of the $x$ dependencies of the
scattered wave amplitudes which are typical for DEAS in the joint arrays
(curves 1 in Fig. 4c, d), into the $x$ dependencies of the scattered wave
amplitudes which are typical for EAS in the uniform arrays (curves 4 in
Fig. 4c, d).

Note that all the curves in Figs. 4a--d are shown only within the region
$0 < x < 2L_1 + L$ (recall that in our examples $L_1 = L_2$). Outside this
region, the relative (normalized) magnitude of the incident wave amplitude
is equal to one (this is the consequence of energy conservation), and the
amplitudes of the scattered waves are the same as at the array boundaries
$x = 0$ and $x = 2L_1 + L$ (i.e. constant).

It can be seen that DEAS can be strong only if the array widths $L_1$ and
$L_2$ (in our examples, $L_1 = L_2$) are less than the critical width $L_c$
that has been determined in our previous
publications~\cite{gramotnev1999b,gramotnev2000}:
\begin{equation}
L_c \approx 2 \left\{ (\mathrm{e}k_1)^{-1}|(E_1)_{x=L_1}[\Gamma_{01}
E_{00}\sin(\eta-\theta_0)]^{-1}|\right\}^{1/2}
\end{equation}
(recall that in the considered examples we have assumed that
$\epsilon_1 = |\epsilon_2|$, i.e. $|\Gamma_{01}| = |\Gamma_{02}|$, and
the critical widths for the first and the second arrays are equal:
$L_c = L_{c1} = L_{c2}$). In Equation (16), $(E_1)_{x=L_1}$ is the maximal
amplitude of the scattered wave at the interface between the two joint
arrays in the limit of large array widths
(i.e. when $L_1 > L_c$)~\cite{gramotnev2000}, and $\mathrm{e} = 2.718$.

Physically, $L_c$ is the distance through which the diffractional divergence
can spread the scattered wave along the $x$-axis inside the array before
this wave is re-scattered by the grating~\cite{gramotnev1999b,gramotnev2000}.
In our previous examples, Equation (16) gives $L_c \approx 15$\,{\textmu}m.
If the width of the arrays $L_1 = L_2 < L_c$
(e.g., $L_1 = 10$\,{\textmu}m---Fig. 4a, c), then for the zero gap width
the diffractional divergence effectively spreads the scattered wave from
the second array throughout the first array, and vice versa. The
interaction between the diffracted waves in both the arrays results
in strong DEAS, i.e. in a resonant increase in the scattered wave
amplitude at a resonant phase shift between the arrays (which should
be relatively close to 180\textdegree)---see \cite{gramotnev1999a,%
gramotnev1999b,gramotnev2000} and curve 1 in Fig. 4c. In this case,
the incident wave amplitude, after an insignificant decrease near the
front boundary, strongly increases, reaching its maximum at the interface
between the arrays---see \cite{gramotnev2000} and curve 1 in Fig. 4a.
The detailed physical explanation of this effect has been presented
in \cite{gramotnev2000}. The scattered wave with a very large amplitude
(typical for the case with $L_1 < L_c$---see curve 1 in Fig. 4c) results
in a strong re-scattered wave that within a very short distance from the
front boundary $x = 0$ appears to be dominating the original incident wave
inside the first array. The amplitude of this re-scattered wave quickly
increases with distance into the first array---curve 1 in Fig. 4a. The
small minimum of curve 1 near the front boundary $x = 0$ is related to
scattering of the incident wave (and thus reducing its amplitude) before
the amplitude of the re-scattered wave becomes dominant (Fig. 4a). Thus,
in the first array, the energy basically flows from the scattered wave
into the incident wave, resulting in a substantial increase in the incident
wave amplitude inside the array. The phase shift between the arrays (that
should be relatively close to 180\textdegree) results in reversing the
situation in the second array. That is, the energy of the powerful (at
the interface $x = L_1$) incident wave starts flowing back into the
scattered wave, resulting in the quick reduction of the incident wave
amplitude back to its vale at the front boundary $x = 0$
(see \cite{gramotnev2000} and curve 1 in Fig. 4a).

If there is a gap between the arrays, then the diffractional divergence
becomes less effective in spreading the scattered wave from one array into
another. This is because the typical gradient of the scattered wave
amplitude across the gap (i.e. across the wave front) decreases with
increasing gap width. Therefore the divergence of the scattered wave from
one array into another becomes weaker. As a result, less effective
interaction of the separated arrays takes place, and the resonant increase
of the scattered wave amplitude becomes smaller (Figs. 2--3). In addition
to this, the minima near the front ($x = 0$) and rear ($x = 2L_1 + L$)
boundaries of the structure become more pronounced with increasing $L$
(Fig. 4a). This is related to the reduction of the scattered wave amplitude
inside the arrays (Fig. 4c), which results in
the reduction of amplitude of the re-scattered wave. Thus the distance
from the front boundary, within which the re-scattered wave becomes dominant
over the incident wave, must increase, i.e. the minimum of the incident wave
near the front boundary must shift towards the middle of the first
array---Fig. 4a. On the left-hand side of this minimum the energy flows
from the incident wave into the scattered wave, and on the other side, in
the opposite direction (up to the rear boundary of the first array).
If $L \ne +\infty$, the overall energy flow in the first array will still
be from the scattered wave into the incident wave (similarly to DEAS in
two joint arrays~\cite{gramotnev2000}). This results in larger amplitudes
of the incident wave in the gap compared to its amplitude at the front
boundary $x = 0$ (Fig. 4a, b). At the same time, if $L\rightarrow +\infty$,
then the amplitude of the incident wave in the gap becomes the same as at
the front boundary. In this case the minimum of the incident wave amplitude
appears to be in the middle of the first array (Fig. 4a), resulting in
overall zero energy flow from the scattered wave into the incident wave.

This pattern of scattering is typical when $L_1 \le L_c$---see Fig. 4a--d.
If the array width $L_1 > L_c$, then the pattern of scattering is different,
and is only weakly dependent on width of the gap between the arrays
(Fig. 5a, b). Moreover, in this case, variations of the scattered and
incident wave amplitudes, caused by varying gap width, are noticeable
only within the regions of width $\approx L_c$ (in our example,
Equation (16) gives $L_c \approx 15$\,{\textmu}m) in each of the arrays
next to the gap (Fig. 5a, b). This is quite expected, since $L_c$
is the distance within which the scattered wave can be spread inside
the array(s) due to the diffractional divergence, before being re-scattered
by the grating~\cite{gramotnev1999a,gramotnev1999b,gramotnev1999c,%
gramotnev2000}.

\begin{figure}[hb]
\includegraphics[width=\columnwidth]{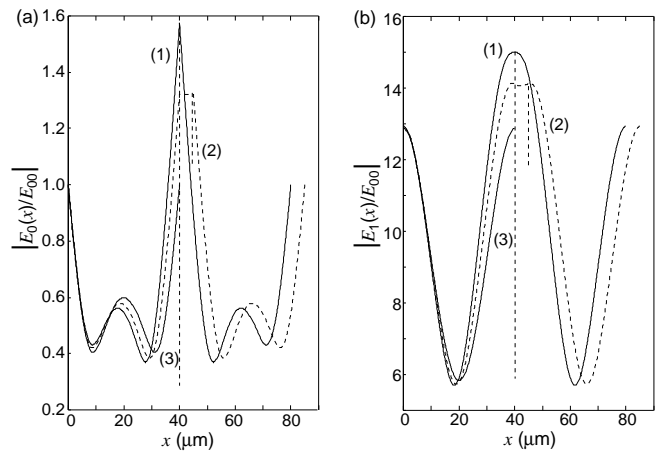}
\caption{Typical dependencies of relative amplitudes of the incident (a)
and scattered (b) waves on distance $x$ from the front boundary $x = 0$
for wide arrays with $L_1 = L_2 = 40$\,{\textmu}m
$> L_c \approx 15$\,{\textmu}m, and for different gap widths: $L = 0$ and
$\phi = \phi_r \approx 158.6$\textdegree\ (curves 1),
$L = 3$\,{\textmu}m and $\phi = \phi_r \approx 153.4$\textdegree\ (curves 2),
$L = +\infty$ (curves 3). The other structural parameters are the same as
for Fig. 4. The edges of the gap between the arrays are marked by the
vertical dotted lines.}
\end{figure}

\section{Applicability conditions of the theory}

Applicability conditions for the two-wave approximation and for the new
approach based on allowance for the diffractional divergence of the
scattered wave are discussed in~\cite{gramotnev2000,nieminen2000}.
In particular, it has been shown that the applicability condition derived
in~\cite{moharam1978,moharam1980,moharam1985} appears to be insufficient
for the case of EAS, and more restrictive inequalities should be used:
\begin{equation}
\begin{array}{l}
4(\lambda_\mathrm{m}/L_1)^2 \ll\\
4(\lambda_\mathrm{m}/L_c)^2 \ll
\end{array}
\begin{array}{l}
\textrm{if}\;\; L_1 \le L_c,\\
\textrm{if}\;\; L_1 > L_c,
\end{array}
\end{equation}
where $\lambda_\mathrm{m}$ is the wavelength in the medium, and $L_c$
is the critical width---see Equation (16).

Use of the two wave approximation (which in the case of strong EAS or
DEAS is basically reduced to neglecting boundary
scattering~\cite{gramotnev2000} will result in a relative error that is of
the order of the left-hand sides of inequalities (17).

Conditions (17) are written for the case of EAS in an isolated uniform
array of width $L_1$ (i.e. for $L = +\infty$ in Fig. 1). For DEAS, the
amplitudes of the scattered wave inside and outside the array can be
significantly larger than for EAS~\cite{gramotnev1999a,gramotnev1999b,%
gramotnev1999c}, and there are four boundaries at which the edge effects
can occur (Fig. 1). However, if $L = 0$ and $L_1 < L_c$, then the two waves
due to boundary scattering from the joint interfaces of the two arrays will
obviously be in antiphase with each other (since the phase shift between the
arrays is about 180\textdegree). The scattered waves due to boundary
scattering from the boundaries $x = 0$ and $x = 2L_1 + L$ will also
interfere destructively. Therefore, the edge effects will mainly result
in an energy flow from the second array to the first array, while the
overall energy flow from the structure will be approximately zero. Since
the arrays are narrower than the critical width $L_c$, the energy flow
between the arrays will be (at least partly) compensated by the
diffractional divergence of the scattered wave.

If there is a gap of width $L \ne 0$ between the arrays, the situation
becomes more complicated, because the waves resulting from boundary
scattering may interfere constructively (depending on gap with $L$).
This may result in a significant energy flow from the structure.
Nevertheless, this energy flow is of the order of the energy flow
between the arrays across the gap.

These speculations show that the approximate theory of DEAS is valid if
the energy flow between the arrays (joint or separated), caused by boundary
scattering, is negligible compared to the energy flow in the incident wave.
Therefore, in accordance with \cite{gramotnev2000}, the condition we are
looking for can be obtained by the multiplication of the left-hand side of
Equation (17a) by the square of the ratio of the scattered wave amplitude
typical for DEAS, $E_1$, to the scattered wave amplitude typical for EAS,
i.e. $(E_1)_{L=+\infty}$:
\begin{equation}
\frac{4\lambda_\mathrm{m}^2}{L_1^2}|(E_1)_{L=+\infty}|^2 \ll 1
\;\;\; \textrm{if}\;\; L_1 \le L_c.
\end{equation}
 
Similarly to inequalities (17), relative errors of using the developed
approximate approach for DEAS in narrow arrays are of the order of the
left-hand side of condition (18).

Finally, the approximate theory of EAS and DEAS, presented in this paper
and in~\cite{bakhturin1995,gramotnev1995,gramotnev1996,gramotnev1997,%
gramotnev1999a,gramotnev1999b,gramotnev1999c,gramotnev2000}, also neglects
the second order derivatives of the incident wave amplitude with respect to
the $x$ coordinate in the array. The analysis of the solutions for the
incident wave inside the array, obtained in~\cite{bakhturin1995,%
gramotnev1995,gramotnev1996,gramotnev1997,gramotnev1999a,gramotnev1999b,%
gramotnev1999c,gramotnev2000}, gives that, despite a significant gradient
of the incident wave amplitude $E_0(x)$, especially in narrow arrays
(see for example curve 4 in Fig. 4a), the second order derivative
$\mathrm{d}^2 E_0/\mathrm{d}x^2$ is about three orders of magnitude
smaller than the term with the first order derivative in Equation (5).
Therefore, with a very good approximation, it can be neglected. Recall
that the allowance for the second order derivative of the scattered wave
amplitude is crucial for the geometry of EAS. In the new approach, this
derivative is automatically taken into account through the consideration
of the diffractional divergence of the scattered wave---see above.

For example, for DEAS described by curves 1--3 in Fig. 4b, d inequality (18)
gives an error less than $\approx 1$\%. However, for curves 1--3 in
Fig. 4a, c, inequality (18) gives errors from $\approx 10$\% for curves
1 to $\approx 3$\% for curves 3. This demonstrates that the two-wave
approximation and the developed approach provide good accuracy for the
considered examples of scattering, especially for the structure with
$L_1 = 15$\,{\textmu}m. Only for very large scattered wave amplitudes
(curves 1 in Fig. 4a, c) is it preferable to use the rigorous
numerical methods~\cite{chu1977,moharam1985,nieminen2000}.

\section{Conclusions}

In this paper we have demonstrated that the diffractional divergence of
the scattered wave in the extremely asymmetrical geometry may result in
a very noticeable interaction of two strip-like periodic arrays separated
by a gap. Due to this divergence, the scattered waves from each of the
separated arrays penetrate into the neighboring array across the gap. As
a result, an optimal (resonant) phase shift between the arrays is shown
to exist, resulting in the double-resonant extremely asymmetrical
scattering, i.e. strong resonant increase of the scattered wave amplitude
in both the arrays and in the gap between them. Clear physical
interpretation of the obtained results, based on allowance for the
diffractional divergence, is presented.

The recently introduced approach for simple analytical (approximate)
analysis of EAS and DEAS, based on the allowance for the diffractional
divergence of the scattered wave, has been extended to the case of two
arrays separated by a gap. One of the significant advantages of this
approach is that it is immediately applicable to all types of waves,
including bulk, guided, and surface optical and acoustic waves in oblique
periodic Bragg arrays with sufficiently small grating amplitude, and
thickness that is much larger than the wavelength of the incident wave
(or the grating period). For example, the derived couple wave equations
describe DEAS of optical modes guided by a slab with a corrugated boundary
if the coupling coefficients $\Gamma_{0j}$ and $\Gamma_{1j}$ are taken
from the known coupled wave theories for the conventional Bragg
scattering~\cite{stegeman1981,popov1985,wellerbrophy1988,hall1990}.
Therefore, all the graphs presented above are valid for slab modes if
the wave numbers of the incident and scattered modes are equal: $k_0 = k_1$
(i.e. there is no mode transformation in the process of scattering), and the
structural parameters of the periodic corrugation are adjusted so that the
numerical values of the coefficients $\Gamma_{0j}$ and $\Gamma_{1j}$
for guided modes are the same as for the considered cases of DEAS
of bulk waves. In addition, the angle of incidence, wavelength of the
slab modes, and the array widths must also be the same as in the
considered examples.

Potential applications of the analyzed effect include signal-processing
devices, optical sensors and measurement techniques (the scattering is
expected to be unusually sensitive to mean structural parameters in the
gap, due to strong dependence of the diffractional divergence on even
small variations of these parameters).

% If in two-column mode, this environment will change to single-column
% format so that long equations can be displayed. Use
%\begin{widetext}
%\end{widetext}

%\bibliography{../../journalsabbrev,../../papers}

\end{document}